\documentclass[aps,prl,twocolumn,superscriptaddress,showpacs]{revtex4}

\usepackage{graphics}

\begin{document}

\title{Collectively optimal routing for congested traffic limited by link capacity}

\author{Bogdan Danila}
\email[]{dbogdan@mail.uh.edu}
\affiliation{Department of Physics, University of Houston, Houston TX 77204-5005}

\author{Yudong Sun}
\affiliation{Department of Physics, University of Houston, Houston TX 77204-5005}

\author{Kevin E.\ Bassler}
\email[]{bassler@uh.edu}
\affiliation{Department of Physics, University of Houston, Houston TX 77204-5005}
\affiliation{Texas Center for Superconductivity, University of Houston, Houston TX 77204-5002}

\begin{abstract}
We show that the capacity of a complex network that models a city street grid 
to support congested traffic can be optimized by using routes that collectively
minimize the maximum ratio of betweenness to capacity in any link. 
Networks with a heterogeneous distribution of link capacities and with a heterogeneous 
transport load are considered. We find that overall traffic congestion and average travel
times can be significantly reduced by a judicious use of slower, smaller capacity links.
\end{abstract}


\maketitle

\section{I. Introduction}
What are the best routes for us to use for driving home tonight in rush hour traffic?
Choosing the best route 
is a task that many of us face everyday.
Knowing the best choice is also important to urban planners who design 
city transportation networks. 
To answer this, and other important questions, a vast amount of research has been devoted 
in recent years to the analysis and optimization of transport on complex 
networks~\cite{BarabScience, Guimera, NewmanSIAM, Helbing, TB, Catanzaro, EcheniquePRE, EcheniqueEPL, Korniss, OurPRE2, YanPRE, OurPRE1, Sreenivasan, OurChaos, OurEPL, Woolf, TKBHK, Barthelemy, Krause-last, Danon, Piontti, Barrat, Motter, Helbing-diagram}. 
If we would collectively choose to use the best routes, then the traffic congestion and 
our commute times could be reduced.

When traveling in a city, the best route to use 
is presumably the one that takes the least time to go from origin to destination. 
If there is little or no traffic, this route is simply the so-called shortest-path route. 
On the complex network that describes the city's streets, 
the shortest-path route is defined as the path 
for which the sum of the weights of each of the links along the path is minimal. 
The weight of a link can be defined as the time it takes to travel from one end of 
the link to the other when traffic is light. Generally, these weights are inversely 
related to the transport capacities of the links (streets that can handle a lot of 
traffic are traveled faster). On the other hand, when traffic is heavy some links 
may become congested, and then the quickest route to the destination may be a longer 
one involving smaller capacity links. It is not obvious, however, exactly at what 
level of congestion the use of alternative routes becomes advantageous, or 
how much the improvements in network transport capacity and average travel time 
amount to. 
The optimal transport routes that we find is the {\it collective optimum} 
that occurs if everyone uses the best routes for collective results. 
As such, it describes an important limit of collective behavior. 
This contrasts with the goal of many traffic studies that seek 
to optimize results individually through a learning process within 
the framework of evolutionary game theory~\cite{Helbing}. 

In this paper, we present methods to help answer these important questions 
and demonstrate the effectiveness of our results 
within the framework of a simple model~\cite{Barthelemy}. 
that captures important characteristics of urban street networks. 
This model has been shown to produce planar geometric networks with an exponential 
distribution of the node distances from the origin and 
with other topological characteristics similar to 
the actual street networks of various cities. 
We use this model to demonstrate the effectiveness of our methods, 
rather than a particular realistic example, in order to make a statistical analysis 
by taking ensemble averages over many example city networks. 
The networks generated using this model have only homogeneous links and are topologically 
more similar to the street networks of older cities rather than those of modern cities. 
To account for the possibility of links with different transport capacities,
we will extend the model to include two types of streets.
Nevertheless, it remains a relatively simple model of street networks, 
and is, thus, ideally suited for the purpose of this paper, which is to develop methods
to determine optimal routes for congested flow. 
Note 
that our results are important not only for the study of 
vehicular traffic but also for other types of link capacity limited network transport, 
including information transport on the Internet. 

Unlike previous studies of 
optimal routing for congested flow~\cite{Guimera, YanPRE, OurPRE1, Sreenivasan, OurChaos, OurEPL}, 
which assumed that traffic is limited by node congestion, here we assume that the capacity
of the network is limited by the amount of traffic each link can support. 
This important and realistic difference requires a nontrivial variation in our methods.
As we have shown previously~\cite{OurPRE1,OurChaos,OurEPL}, when transport is limited 
by node congestion and all nodes have the same processing capacity, 
the transport capacity of the network is maximized by using a set of routes that minimize 
the maximum betweenness of the nodes. 
In~\cite{OurPRE1} we introduced an algorithm for finding such a set of routes and 
demonstrated that, for a number of commonly studied network topologies, 
it finds routes for which the capacity, at least, scales optimally with system size. 
Here we show that when transport is limited by heterogeneous link capacities, 
the transport capacity of the network is maximized by minimizing the maximum 
betweenness to capacity ratio of links. We also consider the traffic optimization problem 
with uneven traffic demands between the various pairs of origin and destination nodes. 
In particular, we study the case of a rush hour traffic burst emanating from 
a central location. 
To obtain our results, we use a variant of our previous routing optimization algorithm. 
Furthermore, we prove a formula that allows the quick computation of the 
average of the sum along the path of any link-related quantities and use 
this formula to compute average travel times.

The rest of the paper is structured as follows. In Sec.~II we present the model we use, 
including details about the network structure, the transport process, and our methods
for determining the optimal routing. In Sec.~III we 
prove a general formula for computing route averages on complex networks in terms 
of link betweennesses and apply that formula to compute the average travel time. In 
Sec.~IV we present our results for network transport capacity and average travel 
time. In Sec.~V we present our conclusions and suggest a few directions for possible
future research.

\section{II. Model}
In our network transport dynamics, particles are assumed to travel along the network 
according to static routing protocols. 
The number of nodes on the network is denoted by $N$. 
We assume that each outgoing link of a node has a separate ``first in, first out" 
(FIFO) particle queue. 
The capacity $C_{ij}$ of link $(i,j)$ is defined as the average number of particles 
transported per time step assuming an infinite number of particles in its queue. 
Transport on the network proceeds in discrete time steps and is driven by inserting new 
particles at the nodes. 
The average number of new particles inserted per time step at node $i$ with destination 
at node $j$ is $r_{ij}$ and we denote $r_i=\sum_{j=1}^N r_{ij}$. 
Each new particle is inserted at the end of the appropriate queue at its node of origin, 
namely the queue corresponding to the first link it has to traverse 
on its way to destination. 
We use a stochastic sequential updating of the positions of particles
that erases the correlations between particle arrival times and ensures that 
both the arrival and the delivery of particles are Poisson processes~\cite{OurChaos}. 
The load of the network is defined as the average $\left<r\right>$ 
over all nodes on the network of $r_i$. The transport capacity of the network
is the critical value of the load $\left<r\right>_c$ above which 
the average particle arrival flux exceeds the processing capacity 
at some link~\cite{Guimera}. 
When this happens, the number of particles in the system will continue to grow and 
the transport becomes {\it jammed}. 

Note that this is not necessarily what happens in the real world, 
where when jamming begins to occur at some link traffic readjusts to avoid the
link.
Nevertheless, our goal here is to study the collective optimum that presumably 
would eventually result from this 
readjustment process. 
As we will see, when the optimal routing finally jams it does 
so simultaneously at a large number of links 
and there is no way to reroute traffic from a jammed link without causing jamming 
at other links.

To account for an uneven traffic demand pattern between the nodes while still using 
a single parameter to characterize the load of the network it is convenient to write 
$r_{ij}=C\left<r\right>\rho_{ij}$, where $\left<r\right>$ is the average number of 
particles per node generated in the course of a time step and 
$\rho_{ij}$ are nonnegative demand weights ($\rho_{ii}=0$). 
If the weights are normalized such that $\sum_{i,j=1}^N \rho_{ij}=N(N-1)$, where $N$ is the number of nodes, we find
\begin{equation}
\label{Eq.1}
    r_{ij}=\frac{\left<r\right>}{N-1}\rho_{ij}.
\end{equation}

\subsection{Network Model}
For simplicity, we consider networks with a binary distribution of link capacities. 
Then, ``streets" are low capacity links and ``highways" are high capacity links. 
Each network realization consists of a network topology, 
generated using the algorithm introduced by 
Barth\'elemy and Flammini~\cite{Barthelemy}, and a set of capacities, 
generated by choosing any link to be a highway with probability $P_h$. 
Our model neglects the correlations that one may expect in real world 
between the capacities of adjacent links. 
These correlations arise from the fact that the subset of links that are ``highways" 
are typically planned to form a well connected subnetwork. 
Enforcing such correlations, however, would have substantially complicated the model. 
In addition, we note that, on one hand, connected sets of highways do form in our model 
and, on the other hand, isolated high capacity links do exist in real street networks. 

The network model we use~\cite{Barthelemy} is based 
on the idea of links growing gradually towards population centers. In its original form, 
the model does not account for any difference in the capacities of the links. 
All links grow at the same constant rate towards the
centroid of their adjacent population centers, 
and new population centers are generated at constant average rate. 
When a link reaches the centroid of its adjacent population centers it splits into 
separate links directed towards each center. 
At the end, the nodes of the network will be the population centers and 
the points where links have been split. 
This model has been shown to generate networks with characteristics similar to those of 
real-world street networks, particularly when the distances of the population centers 
with respect to the city center are exponentially distributed. 
These similar characteristics include the distribution of the node degrees, 
as well as size and shape distributions of the areas delimited by streets. 
Using this model rather than a particular real-world example allows the calculation of 
ensemble averages over thousands of model city networks. 

An ensemble of network realizations is characterized by the number of centers 
generated per time step $R_C$, the rate of growth of the links $d$, 
the total number of centers $N_C$, the ratio between the capacity of a highway and 
that of a street $C_h/C_s$, and the probability $P_h$ for a link to be a highway. 
Note that the number of nodes $N$ on these networks can only be controlled on average, 
through $N_C$ and the ratio $R_C/d$. Without loss of generality, the average distance 
of the population centers from the center of the city is assumed to be unity. 
All results presented below have been computed for $R_C=0.1$, $d=0.001$, and $C_h/C_s=8$. 
Due to the high value of $C_h/C_s$, shortest-path routing will typically forgo 
the use of lower capacity links unless they are strictly needed to achieve connectivity. 

\subsection{Traffic Routing}
The transport capacity of a network depends on the routing protocol used and on 
the processing capacities of the various links. 
To demonstrate the importance of our results, we consider two different types of routing. 
One is the ``natural" shortest path (SP) routing computed with link weights in inverse 
proportion to their respective capacities. 
The other one is the ``optimal" routing (OR) for congested traffic resulting from the 
application of a routing optimization algorithm which finds the set of routes that 
maximizes the transport capacity of the network. 

Note that the problem of finding the absolute best set of optimal routes is 
mathematically related to the problem of finding the minimal sparsity 
vertex separator~\cite{Sreenivasan}, 
which has been shown to be an NP-hard problem~\cite{Bui}. 
In NP-hard problems the number of flops necessary for the computation of an
exact solution can increase with the number of nodes $N$ faster than any polynomial.
Thus, as for other problems that are known to be 
NP-hard~\cite{NewmanSIAM,NewmanPNAS,Radicchi,ArenasNJP,OurEPL86}, it is useful 
to have an algorithm that finds, at least, nearly optimal routes
and runs in only polynomial time. The following algorithm does just that.
Its running time is $O(N^3 \; \log N)$ [$O(N^2 \; \log N)$ for one iteration and requiring 
$O(N)$ iterations].

Routing on a network is characterized by the set of probabilities $P_{ij}^{(t)}$ 
for a particle with destination $t$ currently at node $i$ to be forwarded to node $j$. 
The betweenness $b_i^{(s,t)}$ of a node $i$ with respect to a source node $s$ and 
a destination node $t$ is defined as the sum of the probabilities of all paths 
between $s$ and $t$ that pass through $i$. Node betweenness can be computed in 
terms of the probabilities $P_{ij}^{(t)}$~\cite{Guimera,NewmanPRE}. The betweenness 
of link $(i,j)$ going from $i$ to $j$ is defined by~\cite{NewmanPNAS}
\begin{equation}
\label{Eq.2}
    b_{ij}^{(s,t)}=b_i^{(s,t)} P_{ij}^{(t)}.
\end{equation}
The average number of particles crossing link $(i,j)$ in the course of a time step 
is given by
\begin{equation}
\label{Eq.3}
    w_{ij}=\sum_{s,t=1}^N b_{ij}^{(s,t)} r_{st}.
\end{equation}
Using Eq.~\ref{Eq.1}, the flow of particles through link $(i,j)$ becomes
\begin{equation}
\label{Eq.4}
    w_{ij}=\frac{\left<r\right>B_{ij}}{N-1},
\end{equation}
with the total weighted betweenness of link $(i,j)$ given by
\begin{equation}
\label{Eq.5}
    B_{ij}=\sum_{s,t=1}^N b_{ij}^{(s,t)} \rho_{st}.
\end{equation}

Avoiding jamming requires $w_{ij} \le C_{ij}$ for every link $(i,j)$. 
Consequently, maximum transport capacity is achieved when the highest 
betweenness-to-capacity ratio on the network $(B/C)_{max}$ is minimized and 
is given by
\begin{equation}
\label{Eq.6}
    \left<r\right>_c=(N-1)/(B/C)_{max}.
\end{equation}
The minimization of $(B/C)_{max}$ can be achieved by various methods including, 
but not limited to, simulated annealing and extremal optimization. 
The results presented here were obtained using a variant of 
extremal optimization~\cite{Bak,Boettcher}. 
Specifically, we modified an earlier algorithm that has been shown to converge to 
near-optimal routing in the case of both geometric and small-world networks if traffic 
is limited by the processing capacities of the nodes~\cite{OurPRE1,OurChaos,OurEPL}. 
The original algorithm aims at minimizing the maximum betweenness of any node 
on the network. 
The variant employed here minimizes the maximum betweenness-to-capacity ratio of any link 
on the network. 
It is, therefore, capable of maximizing the transport capacity of networks 
with traffic limited by arbitrary link capacities. The algorithm works as follows:
\begin{enumerate}
\item Assign an initial weight equal to one to every link and compute the shortest paths 
between all pairs of nodes and the betweenness of every link.
\item Find the link that has the highest betweenness-to-capacity ratio $(B/C)_{max}$ 
and increase its weight by one.
\item Recompute the shortest paths and the betweennesses. Go back to step 2.
\end{enumerate}

Note that this algorithm, like many other transport optimization 
algorithms, is essentially a shortest path algorithm with variable weights that 
are tuned in order to reduce congestion. The variable weights allow traffic to
be routed around the locations in the network that are most likely to jam.

\begin{figure}
\begin{center}
	\scalebox{0.4}[0.4]{\includegraphics*{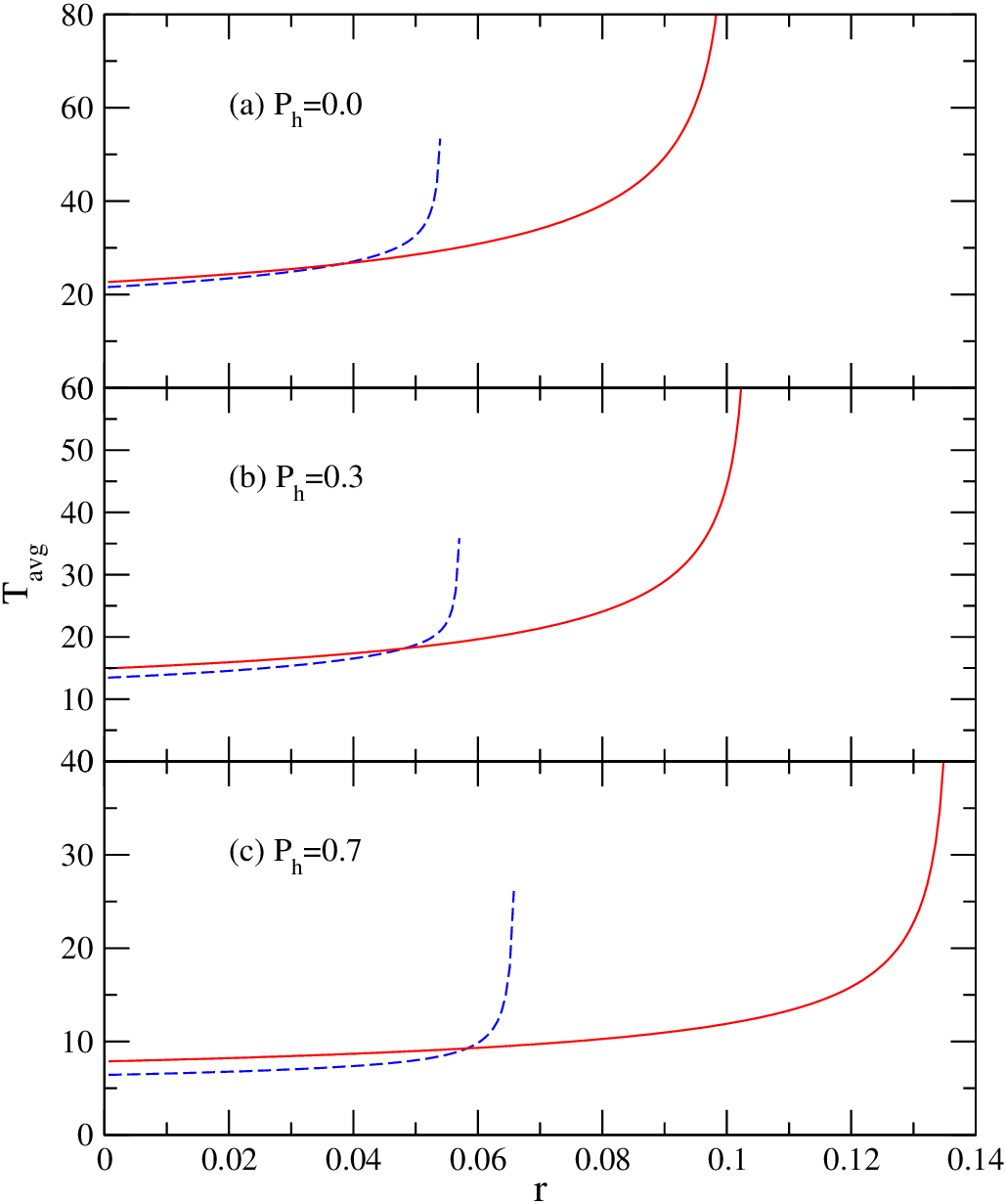}}
	\caption{(Color online) Average travel time (in time steps) vs.\ 
	network load (in particles per time step per node) for 
	a typical network realization with $N=200$ nodes. Results are shown for
	shortest-path routing, SP (blue dashed lines), and for
	optimal routing, OR (red solid lines), at three different
	high capacity link densities, (a) $P_h=0$, (b) $P_h=0.3$, and (c) $P_h=0.7$.}
    \label{Fig.1}
\end{center}
\end{figure}

\begin{figure}
\begin{center}
	\scalebox{0.35}[0.35]{\includegraphics*{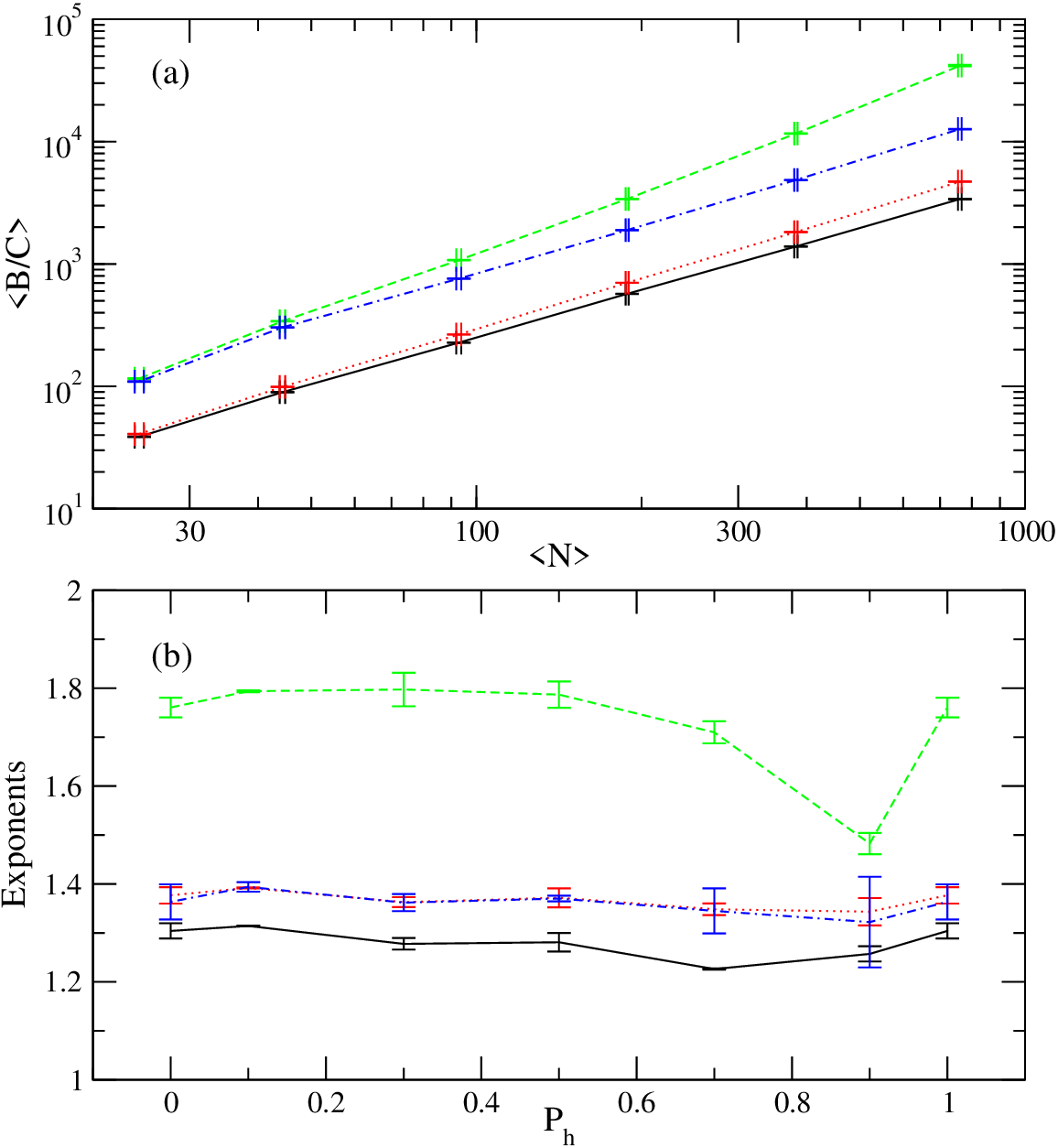}}
    \caption{(Color online) (a) Ensemble averages of the network average and network 
    maximum betweenness-to-capacity ratios, $\left<(B/C)_{avg}\right>$ and 
    $\left<(B/C)_{max}\right>$ respectively, vs.\ average number of nodes at $P_h=0.3$ 
    for networks with an average of 250 nodes. 
    Results are for $\left<(B/C)_{avg}^{SP}\right>$ (black solid), 
    $\left<(B/C)_{avg}^{OR}\right>$ (red dotted), $\left<(B/C)_{max}^{SP}\right>$ 
    (green dashed), and $\left<(B/C)_{max}^{OR}\right>$ (blue dot-dashed). 
    (b) Exponents of the power law fits for the same quantities as functions of $P_h$.}
    \label{Fig.2}
\end{center}
\end{figure}

\section{III. Route averages}
The average travel time can be computed from link betweennesses assuming 
the arrival and the delivery of particles at every node are Poisson processes. 
We will now prove a general formula for calculating averages over 
the entire set of routes considered on a network and then apply this formula to 
calculate the travel time. Let $Q_{ij}$ be any quantity associated 
with the links of a network. To calculate the average over all particle routes 
of the sum of $Q_{ij}$ along the route it is convenient to write the betweenness 
in terms of the probabilities for complete routes. Let $\pi_n(s,t)$ be the ordered 
set of all nodes along the $n$-th distinct route between $s$ and $t$ 
(including $s$ but excluding $t$) and $p_n(s,t)$ be the probability for a particle 
to be routed along that route. The number of distinct routes between $s$ and $t$ 
is $\mathcal{N}(s,t)$. Then,
\begin{equation}
\label{Eq.7}
    b_{ij}^{(s,t)}=\sum_{n=1}^{\mathcal{N}(s,t)} p_n(s,t) \sum_{k\in \pi_n(s,t)} \delta_{ik}\; \delta_{j,next(k|n,s,t)}
\end{equation}
where $next(k|n,s,t)$ is the successor of node $k$ in $\pi_n(s,t)\cup \left\{t\right\}$. 
Let us now compute the quantity $(\Sigma Q)_{avg}^{(s,t)}$ defined by
\begin{equation}
\label{Eq.8}
    (\Sigma Q)_{avg}^{(s,t)}=\sum_{i,j=1}^N Q_{ij} b_{ij}^{(s,t)}.
\end{equation}
By substituting Eq.~\ref{Eq.7} into Eq.~\ref{Eq.8}, we find
\begin{equation}
\label{Eq.9}
    (\Sigma Q)_{avg}^{(s,t)}=\sum_{n=1}^{\mathcal{N}(s,t)} p_n(s,t) \sum_{k\in \pi_n(s,t)} Q_{k,next(k|n,s,t)}.
\end{equation}
The inner sum on the right-hand side of Eq.~\ref{Eq.9} is the sum of $Q_{ij}$ 
along the $n$-th route and $(\Sigma Q)_{avg}^{(s,t)}$ is its average over all routes between $s$ and $t$.

The network-wide average of $(\Sigma Q)_{avg}^{(s,t)}$ weighted by the elements of the traffic demand matrix $r_{st}$ is then
\begin{equation}
\label{Eq.10}
    (\Sigma Q)_{avg}=\frac{1}{N(N-1)}\sum_{i,j=1}^{N} Q_{ij}B_{ij},
\end{equation}
where $B_{ij}$ are defined in Eq.~\ref{Eq.5} and we used the fact that the sum 
of all $\rho_{st}$ equals $N(N-1)$. The derivation above parallels the one presented 
in~\cite{OurChaos}, but the difference in the way one accounts for the links along a 
route in Eqs.~\ref{Eq.7} and~\ref{Eq.9} as opposed to accounting for nodes is non-trivial.

The average time, measured in steps, required for a particle to traverse link $(i,j)$ 
is taken to be
\begin{equation}
\label{Eq.11}
    T_{ij}=\frac{1+\left<q_{ij}\right>}{C_{ij}},
\end{equation}
where $\left<q_{ij}\right>$ is the average length of the queue associated 
with this link and $C_{ij}$ is the link capacity. 
The $1/C_{ij}$ term accounts for the travel time in the limit of low 
traffic and is consistent with the idea that higher capacity links are covered faster. 
The additional travel time due to link congestion is accounted for by the second term. 
Note that this is a very simple model of the traffic flow along the link, 
which we use here for the purpose of illustrating our method. 
More complex (and more accurate) nonlinear empirical formulas could be used to connect 
traffic flow and link capacity to the average travel time along a link. 
Such formulas could also be used in conjunction with Eq.~\ref{Eq.10} to calculate 
network-wide route averages.

Since both the arrival and the delivery of particles at every queue are well approximated 
by Poisson processes, the average queue length is given by~\cite{Allen}
\begin{equation}
\label{Eq.12}
    \left<q_{ij}\right>=\frac{w_{ij}}{C_{ij}-w_{ij}}.
\end{equation}
By using Eqs.~\ref{Eq.11} and~\ref{Eq.12} in~\ref{Eq.10} we obtain the 
average travel time as a function of load
\begin{equation}
\label{Eq.13}
    T_{avg}=\frac{1}{N}\sum_{i,j=1}^{N} \frac{B_{ij}}{(N-1)C_{ij}-\left<r\right>B_{ij}}.
\end{equation}
Additional time delays associated with traveling along the links may also 
be included in the calculation of $T_{avg}$ by using Eq.~\ref{Eq.10}.

\section{IV. Results}
\subsection{Average Travel Time}
Figure~\ref{Fig.1} shows a comparison between the SP and OR average travel times 
as functions of network load $r$ for one network with uniform traffic demand at 
three different values of $P_h$. 
The total travel time along a link is taken to be the waiting time in its queue 
plus an additional delay equal to the inverse of the link's capacity. 
The waiting time in queues accounts for congestion, while the inverse of
a link's capacity accounts for its travel time in the limit of low traffic. 
Note that routing is explicitly optimized to maximize the transport capacity 
and not to minimize the average travel time. 
The results shown are typical for networks with an average 
number of nodes $\left<N\right>\approx 200$. 
This network was generated with a value of $N_C=100$. 
The load above which OR outperforms SP is in this case about 80\% of the critical 
load under SP routing. This may be regarded as the threshold for congestion when 
SP routing is used. 
Optimization increases the transport capacity of this network by a factor of about 1.7 
if all routes have the same capacity, while for values of $P_h$ between 0.2 and 0.8 
the factor of improvement increases to more than 2. 
Note also the decrease of the average travel time at low loads with increasing $P_h$ 
regardless of the type of routing. 
This is due to the increased average capacity of the links.

\begin{figure}
\begin{center}
    \scalebox{0.3}[0.3]{\includegraphics*{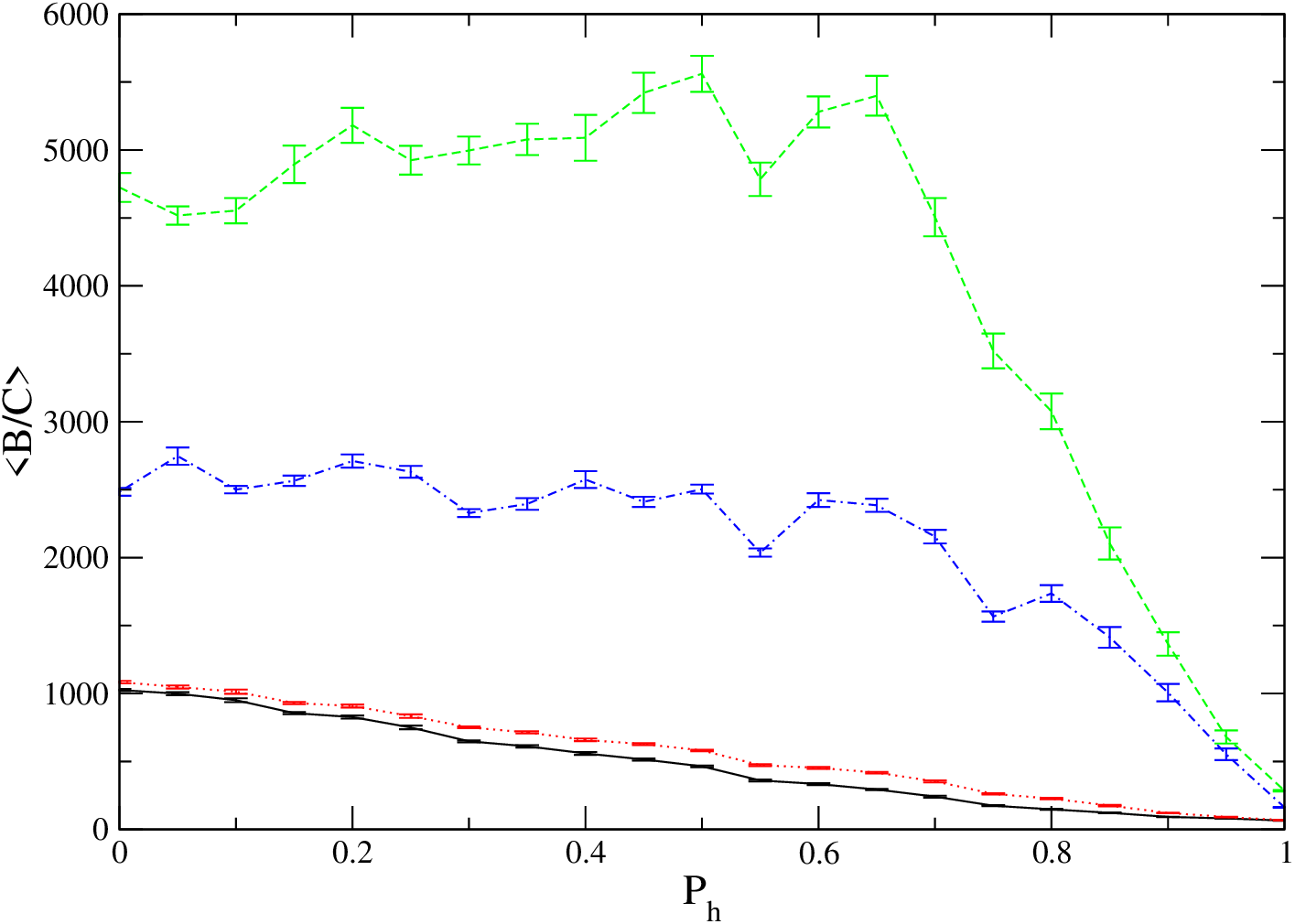}}
    \caption{(Color online) Ensemble averages of the network average and network maximum 
  betweenness-to-capacity ratios, $\left<(B/C)_{avg}\right>$ and $\left<(B/C)_{max}\right>$ 
    respectively, as functions of $P_h$ for networks with an average of 250 nodes. 
    Results are for $\left<(B/C)_{avg}^{SP}\right>$ (black solid), 
    $\left<(B/C)_{avg}^{OR}\right>$ (red dotted), $\left<(B/C)_{max}^{SP}\right>$ 
    (green dashed), and $\left<(B/C)_{max}^{OR}\right>$ (blue dot-dashed).}
    \label{Fig.3}
\end{center}
\end{figure}

\subsection{Network Transport Capacity}
In Fig.~\ref{Fig.2} results are presented for the ensemble averages of 
$(B/C)_{avg}$ and $(B/C)_{max}$ for networks with $\left<N\right>\approx 250$ 
which were generated for $N_C=120$. 
All error bars represent $2\sigma$ estimates. 
The subscripts $avg$ and $max$ denote the average and respectively maximum 
over all links of a network with a given link capacity configuration. 
Averaging over an ensemble of network realizations is denoted by angular brackets. 
This is an average over an ensemble of network topologies and 
over an ensemble of link capacity configurations for each network topology. 
The averages shown in Fig.~\ref{Fig.2} were computed for 100 network 
topologies and 30 link capacity configurations. 
Each ensemble of network realizations is characterized by the average number of nodes 
$\left<N\right>$ and by the probability $P_h$ for a link to be a highway. 

Figure~\ref{Fig.2}(a) shows a log-log plot of $\left<(B/C)_{avg}\right>$ and 
$\left<(B/C)_{max}\right>$ for both SP and OR at $P_h=0.3$. 
Since the lines are nearly straight, the ensemble averages of these quantities 
scale with average network size as a power law. 
Similar power law dependence has been observed in 
other studies~\cite{Sreenivasan, OurChaos} in the case of both the maximum and the average 
node betweenness and is discussed in~\cite{Sreenivasan}. 
Note that any decrease in $\left<(B/C)_{max}\right>$ can only be obtained at the expense of 
an increase in $\left<(B/C)_{avg}\right>$ since avoiding congestion along the shortest path 
means taking longer routes that contribute to the betweenness of more links. 
With OR, the slopes of $\left<(B/C)_{avg}\right>$ and $\left<(B/C)_{max}\right>$ 
are essentially the same.
Thus, the capacity of the routes we find, at least, scales optimally with system size.
This behavior is similar to that of the node betweenness when transport is limited by 
node processing capacity. 
However, the finite size effects are stronger in the current case and the error bars 
under estimate the true error, since they are calculated assuming that the values of 
$(B/C)_{avg}$ and $(B/C)_{max}$ for the various network realizations are 
normally distributed while our simulations show they are not. 
Figure~\ref{Fig.2}(b) shows the exponents of the power law scaling of the four quantities 
as functions of $P_h$. Perhaps somewhat surprisingly, 
note that the exponents for $\left<(B/C)_{avg}^{OR}\right>$ 
and $\left<(B/C)_{max}^{OR}\right>$ are essentially equal over the whole range of $P_h$ 
which, again, argues in favor of the optimality of routing. 
Note also that the exponent for $\left<(B/C)_{max}^{SP}\right>$ 
exhibits a dip around $P_h=0.9$. 
This is an interesting feature, indicating that the SP routing works unusually well 
when there is a small but nonzero concentration of low capacity links. 

Figure~\ref{Fig.3} shows $\left<(B/C)\right>$ versus $P_h$ for $\left<N\right>\approx 250$ 
generated for $N_C=120$. 
The averages shown in this figure were also computed for 100 network 
topologies and 30 link capacity configurations. 
The error bars represent $2\sigma$ estimates. 
Note that $\left<(B/C)_{avg}\right>$ varies monotonically between $P_h=0$ and 
$P_h=1$ corresponding to all streets and all highways, respectively, 
while $\left<(B/C)_{max}\right>$ exhibits a midrange maximum in the case of SP routing 
that corresponds to a dip of the network transport capacity. 
This type of behavior is due to the fact that SP routing forgoes the use of low capacity 
links as long as they are not strictly needed to achieve connectivity (just as we tend to 
do in real life), which increases congestion on highways. On the other hand, by optimally 
using all links, it is possible to avoid this phenomenon.

\begin{figure}
\begin{center}
    \scalebox{0.4}[0.4]{\includegraphics*{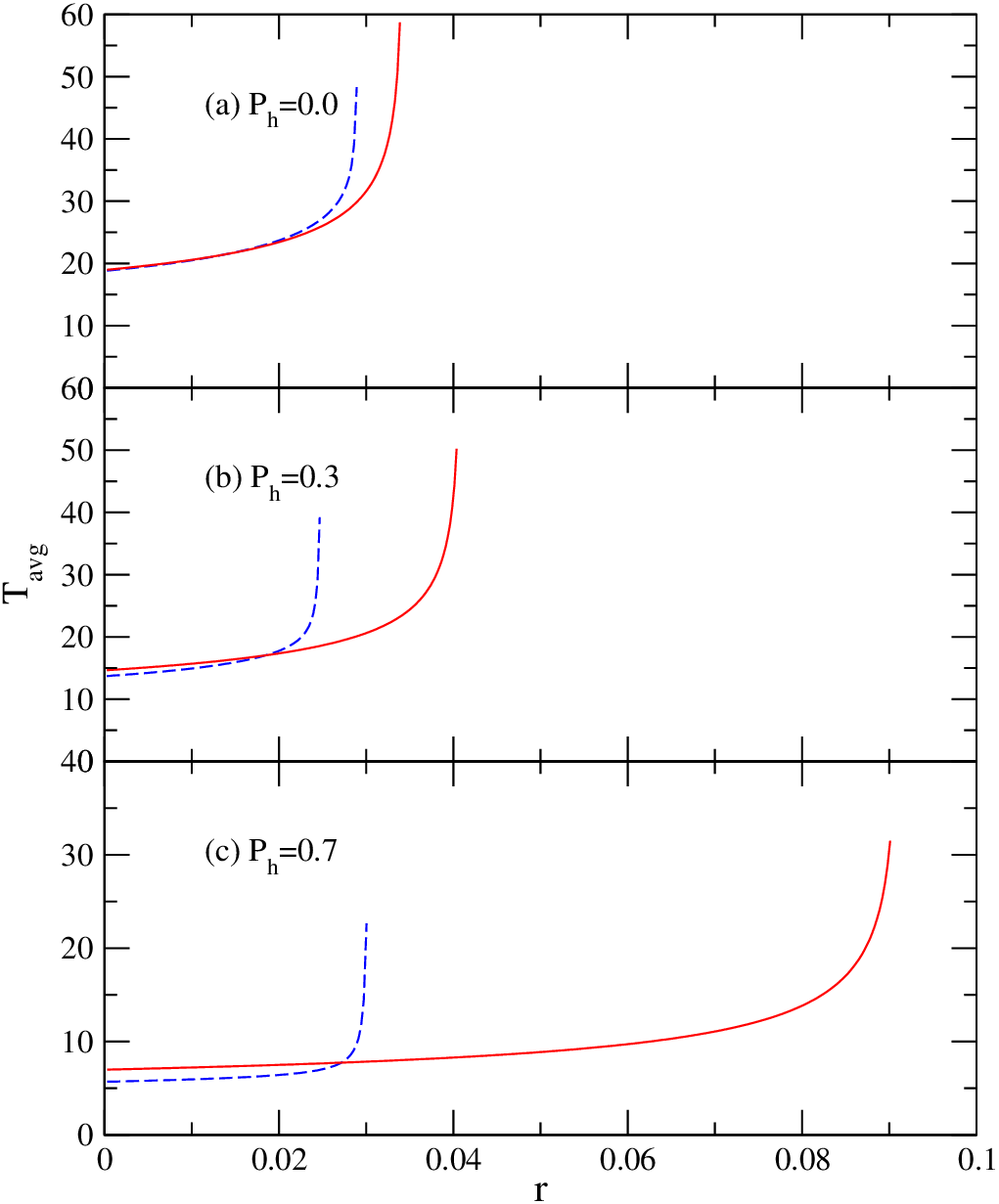}}
    \caption{(Color online) Average travel time (in time steps) vs.\ network load 
    (in particles per time step per node) for a ``rush hour traffic burst" 
    with particles originating from the innermost 10\% of the nodes 
    on a typical network realization with $N=200$ nodes. 
    Results are shown for shortest-path routing, SP (blue dashed lines), and for optimal 
    routing, OR (red solid lines), at three different high capacity link densities, 
    (a) $P_h=0$, (b) $P_h=0.3$, and (c) $P_h=0.7$.}
    \label{Fig.4}
\end{center}
\end{figure}

\subsection{Uneven Traffic Demand}
A comparison between the SP and OR average travel times for one network with uneven traffic 
demand is presented in Fig.~\ref{Fig.4}. Specifically, we look at a ``rush hour 
traffic burst" with particles originating from the innermost 10\% of the nodes that
we use to model a ``downtown". 
Their destinations are chosen uniformly from among all nodes with the only constraint being 
that the node of origin must be different from the destination. 
This is the same network that was used for Fig.~\ref{Fig.1} and again the results are 
representative for networks with an average of about 200 nodes. 
Note the lower maximum capacity of the network, 
which is due to the uneven distribution of traffic. Nevertheless, a judicious use of 
the low capacity links again results in a significant increase of the transport capacity. 
The factor by which transport capacity is increased may be even higher than 
in the case of uniform traffic demand.
This result is particularly important since traffic congestion usually develops is 
situations of uneven traffic demand.

\section{V. Conclusions}
In summary, we have presented a quantitative study of collective 
transport optimization on planar networks that model a city street grid. 
Traffic on these networks is limited by link congestion. 
The network model accounts for many topological features seen in real life urban 
street networks and allows for heterogeneous link capacities and uneven traffic demands. 
The question posed in the title is answered in statistical terms, as we show that a 
judicious use of all links, including low capacity ones, can significantly alleviate 
traffic congestion. 
The results we present show that the average factor by which transport capacity can be 
increased varies with network size as a power law, and that the routes we 
found are ones that allow the capacity to, at least, scale optimally with system size.
Moreover, the average travel times on congested networks can be significantly decreased. 
We determine the network load at which using the low capacity links becomes advantageous 
from the point of view of the average travel time. 
Most importantly, we show that significant improvements in traffic routing can be achieved 
not only in the case of uniform traffic demand, 
but also in situations of extremely uneven traffic. 
Our results are important not only for quantifying the routing improvements that 
can be achieved on existing networks, but also for the design of future street networks. 
For example, numerical experimentation with optimal traffic routing on a model of a real 
city street network can pinpoint the links for which an increase in capacity will 
result in the highest network transport capacity. 

It would be interesting to extend this study to 
empirical city road networks, and to
more realistic models that would include the correlations between the 
capacities of adjacent links corresponding to highway structures. 
A more realistic model could also incorporate traffic fluctuations~\cite{Motter}. 
In addition, recent work~\cite{Helbing-diagram} has led to better understanding of the 
observed nonlinear relationship between the average travel time on a link and 
the load to capacity ratio. 
A more theoretical understanding of why the power law behavior of 
$\left<(B/C)_{max}\right>$ and $\left<(B/C)_{avg}\right>$ is exhibited even by 
networks that are not scale-free and of the relative independence of the exponents on 
$P_h$ are also interesting and important open questions.

\begin{acknowledgments}
This work was supported by the NSF through grants Nos.\ DMR-0427538 and DMR-0908286
and by the Texas Advanced Research Program through grant No.\ 95921.
\end{acknowledgments}

\end{document}